\documentclass[twocolumn]{elsart}

\usepackage{graphics}

\usepackage{graphicx}
\usepackage{epsfig}

\usepackage{amssymb}
\usepackage{amsmath}

\begin{document}

\begin{frontmatter}



\title{An Ion Guide for the Production of a Low Energy Ion Beam of Daughter Products of $\alpha$-Emitters}

\author[manchester]{B. Tordoff \corauthref{cor1}}\ead{bwt@phys.jyu.fi},
\author[Finland]{T. Eronen},
\author[Finland]{V.V. Elomaa},
\author[Canada]{S. Gulick},
\author[Finland]{U. Hager},
\author[Finland]{P. Karvonen},
\author[Finland]{T. Kessler},
\author[Canada]{J. Lee},
\author[Finland]{I. Moore},
\author[Russia]{A. Popov},
\author[Finland]{S. Rahaman},
\author[Finland]{S. Rinta-Antila},
\author[Finland]{T. Sonoda},
\author[Finland]{J. \"Ayst\"o},

\corauth[cor1]{Tel.: +358-14-260-2440; fax:+358-14-260-2351 }

\address[manchester]{Nuclear Physics Group, Schuster Laboratory, Brunswick Street, University of Manchester, Manchester, M13 9PL, UK}
\address[Finland]{Department of Physics, University of Jyv\"askyl\"a, P.O. Box 35, 40014, Jyv\"askyl\"a, Finland }
\address[Canada]{Ernest Rutherford Physics Building, McGill University, 3600 rue University, Montr$\acute{e}$al, QC, H3A 2T8  Canada }
\address[Russia]{Petersburg Nuclear Physics Institute, Gatchina, St-Petersburg, 188350, Russia}

\begin{abstract}
A new ion guide has been modeled and tested for the production of a low energy ($\approx$ 40 kV) ion beam of daughter products of alpha-emitting isotopes. The guide is designed to evacuate daughter recoils originating from the $\alpha$-decay of a $^{233}$U source. The source is electroplated onto stainless steel strips and mounted along the inner walls of an ion guide chamber. A combination of electric fields and helium gas flow transport the ions through an exit hole for injection into a mass separator. Ion guide efficiencies for the extraction of $^{229}$Th$^{+}$ (0.06\%), $^{221}$Fr$^{+}$ (6\%), and $^{217}$At$^{+}$ (6\%) beams have been measured. A detailed study of the electric field and gas flow influence on the ion guide efficiency is described for two differing electric field configurations.  
\end{abstract}

\end{frontmatter}

% main text
\section{Introduction}
\label{intro}
\noindent In order to probe the structure of low energy nuclear isomeric states whose surroundings can substantially influence the properties of the state \cite{dykhne}, it has become necessary to develop a means of performing measurements in a medium-free environment. The method described here implements the Ion-Guide Isotope Separator On-line technique conceived \cite{valli}, developed \cite{arje,dend,Juha,Lait} and successfully used at the University of Jyv\"askyl\"a for over 20 years. In the technique, nuclear reaction products are thermalised in fast flowing helium gas and are extracted as ions through an exit hole for injection into a mass separator. The Jyv\"askyl\"a IGISOL system has recently been upgraded, both in general design \cite{Heikki} in order to handle higher primary beam currents and in post-gas cell ion guidance. This latter upgrade has resulted in the installation of a radio-frequency sextupole ion guide which replaces the skimmer system based on previous tests in Jyv\"askyl\"a \cite{Jussi} and elsewhere \cite{Xu,vanden}. The rf-sextupole has the advantage of a higher overall mass resolution and better transmission efficiency. A laser ion source has also recently been developed \cite{iain} based upon resonance ionisation in a gas cell \cite{vermeeren,facina} and variations thereof \cite{blaum}. Developments in ion guide technology \cite{Jussi} have shown that the addition of in-guide electric fields improves the extraction efficiency and reduces the evacuation time of radioactive ions. 

\noindent In this work two electric field arrangements have been designed and tested for the evacuation of daughter recoils originating from a $^{233}$U source inside a large volume gas cell. Firstly, a standard electrostatic field has been modeled and tested employing charged stainless steel plates close to the guide exit hole based upon an earlier design \cite{arto1}. Secondly, an electron emitter design \cite{Jussi} has been tested, whereby a potential difference is created in the guide via a source of electrons close to the exit hole.

% \appendix
%%%%%%%%%%%%%%%%%%%%%%%%%%%%%%%%%%%%%%%%%%%%%%%%%%%%%%%%%%%%%%%%%%%%%%%%%%%%%%%%%%%%%%%%%%%%%%%%%%%%%%%%%%%%%%%%%%%%%%%%%%%%%%%%%%%%%%%%%%%%%%%%%%%%%%%%%%%%%%%%%%%%%%%

\section{Ion Guide Design}

\noindent The ion guide dimensions are specifically designed to maximise the production of a radioactive ion beam of $^{229}$Th dictated by the shape and $\alpha$-recoil energy of the source. Moreover, the guide is adaptable to any $\alpha$-recoil source and is fully compatible with both the new laser ion source and rf-sextupole ion guide. 

\subsection{The $^{229}$Th$^{+}$ source}
\label{thsource}
\noindent The dimensions of the ion guide were chosen so that the maximum range of 84 keV $\alpha$-recoils in 50 mbar He  gas (10 mm) was smaller than the guide radius (57.5 mm). The source of $^{229}$Th was produced (Eq. \ref{beta}) through neutron capture and subsequent $\beta^{-}$ decay of unstable nuclei from a seed nucleus of $^{232}$Th. This source was supplied by Isotope Products Laboratory, Bubank, USA. 

\begin{equation}
^{232}Th(n,\gamma )^{233}Th\stackrel{\beta^{-}}{\rightarrow}\:  ^{233}Pa\stackrel{\beta^{-}}{\rightarrow} \: ^{233}U
\label{beta}
\end{equation}

\noindent The $^{233}$U parent was dissolved in isopropanol and chemically electroplated onto 12 stainless steel sheets with dimensions 76.2 mm $\times$ 25.4 mm. These strips were then mounted on the inside surface of the gas cell defining its diameter to be 115 mm. 

\begin{figure}
	\centering
		\includegraphics[width=\linewidth]{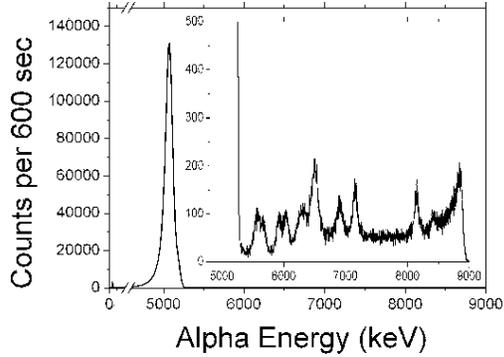}
	\caption{Direct alpha spectrum from one $^{233}$U foil taken over a 10 minute collection time.}
	\label{fig:figure1}
\end{figure}

\noindent The strength of the source supplied was measured through analysis of its $\alpha$-spectrum (Fig. \ref{fig:figure1}). The dominant $\alpha$-decay peak arising from the decay of $^{233}$U to $^{229}$Th was fitted yielding a source strength of 10$^{6}$ Bq. The inset of Fig. \ref{fig:figure1} shows a magnified portion of the spectrum where alpha peaks relating to decays lower in the decay chain have been identified. Table \ref{tab:sourcea} lists the alpha energy and corresponding isotope of these peaks. The presence of decays from the $^{232}$Th chain indicates a population of impurities not fully suppressed in the extraction of $^{233}$U from the breeder reactor.  

\begin{table}[htbp]
	\centering
		\begin{tabular}{||c|c|c||}
\hline Isotope&Source&Alpha Energy(MeV)\\\hline
$^{229}$Th& $^{233}$U & 4.845, 4.814, 4.839 \\\hline
$^{224}$Ra&$^{232}$Th& 5.678 \\\hline
$^{225}$Ac&$^{233}$U& 5.817 \\\hline
$^{212}$Bi& $^{232}$Th & 6.089, 6.050 \\\hline	
$^{221}$Fr& $^{233}$U & 6.260  \\\hline
$^{220}$Rn&$^{232}$Th&  6.278 \\\hline
$^{216}$Po& $^{232}$Th & 6.778  \\\hline	
$^{217}$At& $^{233}$U & 7.066 \\\hline			
$^{213}$Po& $^{233}$U & 8.375 \\\hline	
$^{212}$Po& $^{232}$Th & 8.784 \\\hline	
		\end{tabular}
 \caption{Alpha peak energies and origin in the source $\alpha$-spectrum.}
	\label{tab:sourcea}
\end{table}

\subsection{Gas Flow Ion Guidance}

\begin{figure}
	\centering
		\includegraphics[width=\linewidth]{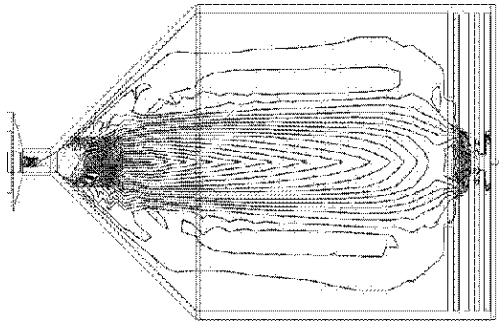}
	\caption{Helium gas flow velocity distribution for an ion guide pressure of 50 mbar. Each isovelocity contour denotes 0.4 ms$^{-1}$ of gas flow. Gas enters from the left and exits through the hole on the right.}
	\label{fig:figure2}
\end{figure}

\noindent The helium gas flow was numerically modeled \cite{cosmo} for the ion guide and results can be seen in Fig. \ref{fig:figure2}. The injection of He buffer gas on the symmetry axis of the guide creates a localised fast flowing region, whose velocity stagnates at larger guide radii. Increasing the pressure inside the guide to 50 mbar confines the region to a smaller volume and increases the flow speed to 50 ms$^{-1}$ on the symmetry axis. To take advantage of this flow, electric field configurations have been designed which preferentially direct charged recoils from the walls of the chamber to the central fast flowing gas region. 

\noindent The evacuation time of the entire guide due to gas flow alone can be estimated from the cell volume and exit hole conductance. The cell evacuation time, $T$ [s], is empirically approximated for a helium filled cylinder by \cite{arje2}

\begin{equation}
T = \frac{V }{C} \approx \frac{\pi r^{2}z}{4.51\times10^{5} \phi_{exit}^{2}} = 690 \; \text{ms}
\end{equation}

\noindent where $V$ is the cell volume, $C$ [m$^{3}$s$^{-1}$] is the hole conductance, $\phi_{exit}$ is the exit hole diameter (2 mm) and $r$ (57.5 mm) and $z$ (120 mm) are guide dimensions. The constant in the denominator (4.51$\times$10$^{5}$) is in units of [mm s$^{-1}$] and is dependent upon the elemental composition of the buffer gas. One can estimate the mean lifetime, $\tau$ [s], of an ion diffusing from the symmetry axis  of an infinitely long cylinder of radius $r_{0}$ [mm] and pressure $p$ [mbar] at room temperature as \cite{arje2}

\begin{equation}
\tau = \frac{p}{500}\left(\frac{r_{0}}{24.05}\right)^{2}.
\end{equation}

\noindent Since this corresponds to 570 ms at 50 mbar with a cylinder radius of 57.5 mm, diffusion to the walls occurs on a comparable timescale to the total guide evacuation time if gas flow alone is considered.

\subsection{Modeling of Electrostatic Field Guidance}

\begin{figure}
	\centering
		\includegraphics[width=\linewidth]{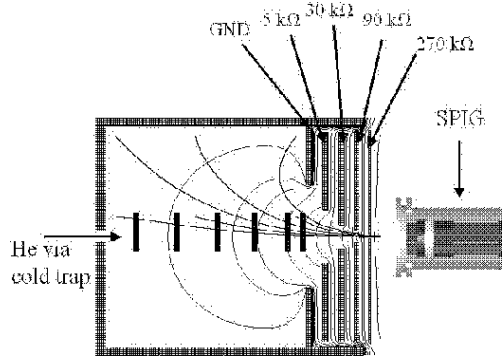}
	\caption{Schematic drawing showing the positions of the $\alpha$- recoil source for determining the ion guide efficiency.}
	\label{fig:figure3}
\end{figure}

\noindent The extraction end of the gas cell is adaptable to two electric field arrangements. Firstly, a series of charged stainless steel plates of decreasing inner diameter (Table \ref{tab:diameter}) create an electrostatic field which penetrates into the gas cell and draws charged $\alpha$-recoils to the exit hole region (Fig. \ref{fig:figure3}).  

\begin{table}[htbp]
	\centering
		\begin{tabular}[width=\linewidth]{||p{1.5cm} |p{1cm} |p{1cm} |p{1.6cm} ||}
\hline Electrode Number& Inner $\phi$ (mm)& Outer $\phi$ (mm) & Resistance (k$\Omega$)\\\hline
1&50&130&0 \\\hline
2&26&95&5 		\\\hline
3&14&95&30  \\\hline	
4&6&95&90 \\\hline	
5&2&115&270	\\\hline
		\end{tabular}
	\caption{Electrostatic ring dimensions. The electrodes are numbered sequentially from the gas cell toward the exit hole.}
	\label{tab:diameter}
\end{table}

\begin{figure}
	\centering
		\includegraphics[width=\linewidth]{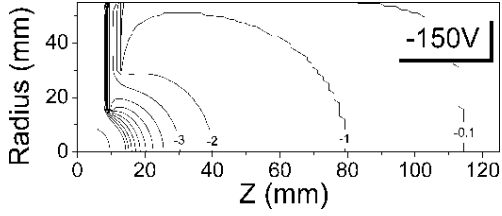}
	\caption{Equipotential lines arising from charged stainless steel plates between the ion guide exit hole and gas cell.}
	\label{fig:figure4}
\end{figure}

\noindent The shape and distribution of equipotential lines within the guide are shown in Fig. \ref{fig:figure4} for an applied voltage of -150 V placed on the outermost electrode (electrode 5) in the resistor chain. The field gradient is variable through interchanging resistance values connecting each plate. Optimum values for the resistance ratios were deduced through modeling of the ion guide by the program SIMION \cite{simion}. Care was taken to maximise the penetration depth of the electric field and to provide a symmetric field which would guide the ions to the exit hole of the guide. 

\begin{figure}
	\centering
		\includegraphics[width=\linewidth]{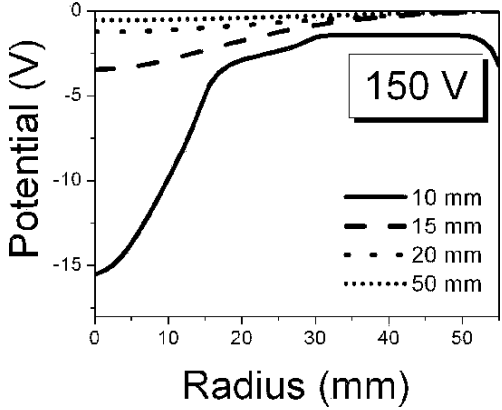}
	\caption{Potential as a function of radius for various distances from the exit hole of the electrostatic plate guide. Increasing the voltage on the electrostatic plates only provides a significant potential close to the guide exit hole. }
	\label{fig:figure5}
\end{figure}

\noindent The potential as a function of radius for different distances from the exit hole was modeled (Fig. \ref{fig:figure5}) indicating that distances further than 2 cm from the the hole are effectively redundant due to the weak potential difference present in this region. Ions close to the exit hole will be successfully extracted, however some fraction of ions will be lost due to collisions with the guide front wall. This problem has been addressed successfully elsewhere \cite{Morita,wada} through the use of rf electric fields to prevent ions hitting the guide walls. 

\subsection{Modeling of Electron Emitter Field Guidance}

\begin{figure}
	\centering
		\includegraphics[width=\linewidth]{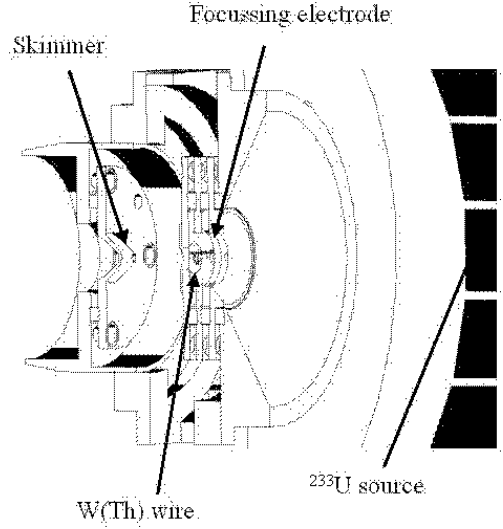}
	\caption{Cut through view of the electron emitter end plate design. The gas cell is on the right of the picture and the mass separator is to the left.}
	\label{fig:figure6}
\end{figure}

\begin{figure}
	\centering
		\includegraphics[width=\linewidth]{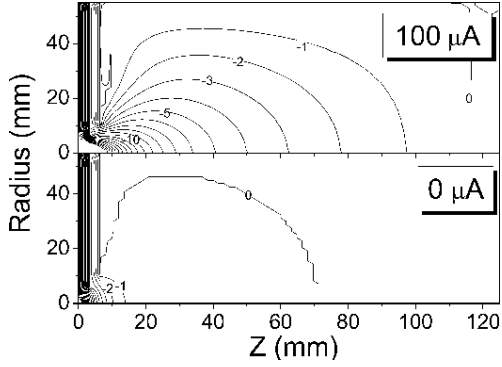}
	\caption{In-guide equipotential lines arising from an electron emitting W(Th) wire surrounding the guide exit hole. Emitter currents of 100 $\mu$A (top) and 0 $\mu$A (bottom) are shown in the figure for an ion guide pressure of 50 mbar.}
	\label{fig:figure7}
\end{figure}

\noindent A second variation of the ion guide extraction region includes a circular focussing electrode followed by a loop of thoriated tungsten W(Th) wire (Fig. \ref{fig:figure6}). The wire is doped with 1\% $^{232}$Th in order to reduce the work function. By passing an electric current through the wire, it is Joule heated to $\approx$ 2000 K and electron emission occurs. This creates a potential difference inside the guide due to the spatial distribution of electrons which draws positively charged $\alpha$-recoils through the exit hole. By solving Poisson's equation iteratively using the particle-in-cell technique, the resulting electric field has been modeled for various emitter currents, an example of which can be seen in Fig. \ref{fig:figure7}.

\begin{figure}
	\centering
		\includegraphics[width=\linewidth]{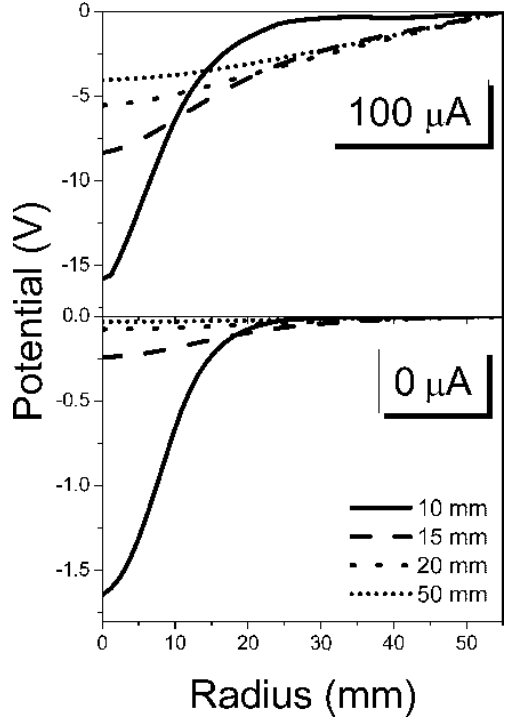}
	\caption{Potential as a function of ion guide radius for various distance from the exit hole of the electron emitter guide. As the electron emitter current is increased, the potential gradient within the guide increases providing more efficient and faster ion extraction.}
	\label{fig:figure8}
\end{figure}

\noindent The potential as a function of ion guide radius can be seen in Fig. \ref{fig:figure8} for the case of the electron emitter, showing a greater potential difference further from the exit hole compared to the electrostatic plate potential (Fig. \ref{fig:figure5}). The potential gradient with no emitter current arises from a focussing electrode surrounding the exit hole and a skimmer electrode placed after the W(Th) wire. 

\section{Ion Guide Efficiency Tests}
\label{tests}
\noindent In order to determine the extraction efficiency of the ion guide with the two different electric field arrangements, an $\alpha$-recoil source of  $^{223}$Ra \cite{Jussi} was collected onto a 3 cm diameter metal coin fixed to a rigid aluminium rod. 

Alpha recoils of its decay product, $^{219}$Rn, were ejected from the source and guided to the exit hole by a combination of the in-guide electric fields and helium gas flow. The ions were guided through the rf-sextupole, mass separated and implanted onto a foil in front of a silicon detector in the focal plane of a 55$^{o}$ mass analysing magnet. Alpha decays from the daughter of $^{219}$Rn,  $^{215}$Po, were detected yielding an ion guide efficiency, relative to the total number of $\alpha$-recoils emanating from the source. The ion guide efficiency was studied as a function of $\alpha$-source distance from the exit hole for the two distinct in-guide electric field configurations and includes a 30\% transmission efficiency from the ion guide to the focal plane of the magnet.

\subsection{Efficiency of Extraction of $\alpha$-Recoils Using Electrostatic Field Guidance}

\begin{figure}
	\centering
		\includegraphics[width=\linewidth]{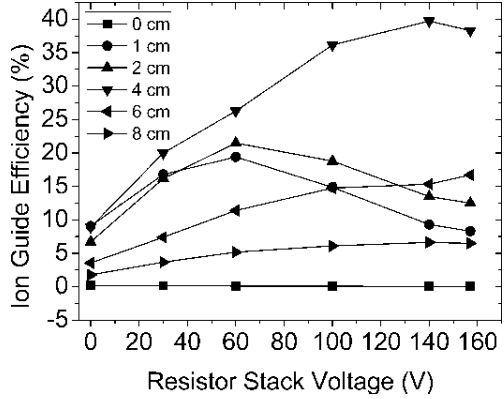}
	\caption{Ion guide efficiency as a function of resistor stack voltage for different distances from the exit hole.}
	\label{fig:figure9}
\end{figure}

\noindent With an ion guide pressure fixed at 50 mbar to provide buffer gas thermalisation, the efficiency as a function of the fifth electrode voltage was measured at various distances from the guide exit hole, along the symmetry axis (Fig.\ref{fig:figure3}). The results are shown in Fig. \ref{fig:figure9}. At the zero position an efficiency close to 0\% is observed arising from the range of recoils being larger than the source - wall distance. At short distances from the exit hole, the gas flow and electric field are perturbed by the presence of the $\alpha$-recoil source providing somewhat misleading results. The efficiency then increases to a maximum value at 4 cm from the exit hole. It is reduced at larger distances due to a compromised range of ions, even in a guiding electric field. 

\subsection{Efficiency of Extraction of $\alpha$-Recoils Using Electron Emitter Field Guidance} 

\begin{figure}
	\centering
		\includegraphics[width=\linewidth]{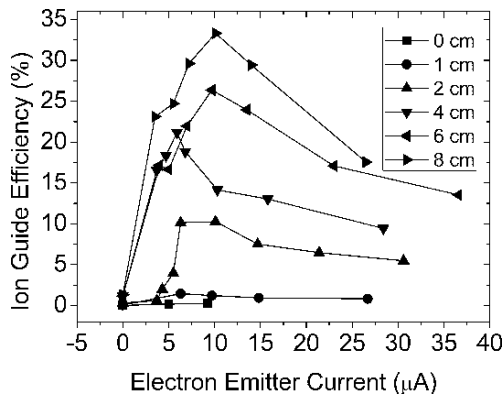}
	\caption{Ion guide efficiency as a function of electron emitter current for a fixed pressure of 50 mbar.}
	\label{fig:figure10}
\end{figure}

\noindent The electron emitter guide does not require the use of the SPIG as it has its own skimmer system which connects directly to the front end of the beamline. The effect of an electron emitter current upon the ion extraction efficiency has been investigated previously \cite{Jussi} for small gas cell radii ($r$ = 30 mm). This work extends the applicability of the method to larger gas cell radii ($r$ = 57.5 mm) and extended alpha source spatial distributions as well as off-axis geometries. The electron emitter current required to provide a given space charge effect is a function of the electron mobility and therefore helium gas pressure. Helium gas pressure also dictates the temperature of the emitter wire and the corresponding amount of electron emission. At a fixed helium pressure of 50 mbar large space charge effects can be observed with relatively low emitter currents. Fig. \ref{fig:figure10} shows the ion guide efficiency as a function of electron emitter current with the $\alpha$-recoil source at various positions along the axis of the ion guide. For recessed positions inside the guide relative to the exit hole, the efficiency increases rapidly as a function of electron emitter current. The efficiency peaks at approximately 10 $\mu$A for all source positions. This is due to a saturation of electrons within the guide at this value. The efficiency drops at higher emitter currents, but the processes leading to this trend are not fully understood.

\noindent In comparison with the electrostatic guide, the maximum peak efficiency obtained is similar, however the integrated total guide efficiency is an order of magnitude higher. This can be attributed to a greater potential gradient throughout the guide, which is particularly evident at large distances from the exit hole where the electron emitter potential gradient is an order of magnitude greater than that of the electrostatic potential. A further striking difference between the two methods is that the maximum efficiency increases as a function of distance for all positions in the electron emitter guide whereas this is not the case for the electrostatic guide.

\section{Extraction of a Thorium Ion Beam}

\subsection{Production Efficiency of Mass Separated Ion Beams}

\noindent Based upon the results in section \ref{tests}, the electron emitter guide system was chosen to provide electric field guidance for production of the thorium ion beam. The pressure was again fixed at 50 mbar and the electron emitter current was maintained at 5 $\mu$A which were experimentally determined optimum values for the off-axis $^{229}$Th source.

\noindent Mass separated alpha spectra arising from decays lower in the $^{233}$U decay chain were taken and alpha peaks attributed to $^{221}$Fr (t$_{\frac{1}{2}}$= 4.9 min) and $^{217}$At (t$_{\frac{1}{2}}$= 32.3 ms) were observed. Both rates were of the order of 30 s$^{-1}$  which, through calibration of the source strength, results in a total ion guide efficiency of 6\% $\pm$ 0.5\% for the extraction of both $^{221}$Fr and $^{217}$At. Since the population of both francium and astatine are in equilibrium in the source, the fact that both are extracted with the same efficiency indicates that the evacuation time of the guide is less than the half life of $^{217}$At, otherwise it would have a lower yield. If one compares this lifetime to the 680 ms total guide evacuation time given simply by the cell volume and hole conductance, it is clear that the electric field provides at least an order of magnitude reduction in extraction time compared to gas flow alone. 

\subsection{Ion Beam Identification with the JYFLTRAP Double Penning Trap}

\noindent As the half life of $^{229}$Th is 7880 years \cite{goldstein}, particle identification was carried out by isobaric mass separation using the JYFLTRAP Penning trap \cite{kolhinen}. The Penning trap is situated after a rf-quadrupole cooler \cite{arto} and is segmented into two traps, one for beam purification and the second for precision mass measurements. In this work only the purification trap was used and particle identification was performed through a mass independent dipole excitation followed by a mass selective ($\frac{M}{\Delta M}$ = 23500 for $^{229}$Th$^{+}$) quadrupole excitation, centering ions of interest onto the beamline axis for single ion detection. The purification trap is buffer gas filled (10$^{-4}$ mbar) to provide cooling of ions during the quadrupole excitation. Bunches are formed inside the rf-quadrupole cooler prior to injection into the purification trap. A bunching time of 660 ms was employed, corresponding to the optimum mass resolution settings for the mass of interest. A transmission efficiency of 10\% was measured from the switchyard to the ion detector after the Penning trap by comparison of the $^{221}$Fr switchyard alpha activity and post trap ion rate respectively.\\

\noindent The relative abundances of molecules formed in the ion guide was measured (Table \ref{tab:mols}) by comparing the integrated number of counts in the quadrupole excitation resonance of a given molecule to that of bare $^{229}$Th$^{+}$ over the same cycle time. A singly charged $^{229}$Th ion can only be formed inside the ion guide either by direct emission in such a state from the source or via three body interactions with impurities in the gas from a higher charge state. It cannot charge exchange with helium due to the first, second and third ionisation potentials of thorium being less than the first ionisation potential of helium. Given this required process, molecular sidebands are to be expected in the mass spectrum of the source.  

\begin{table}[htbp]
	\centering
		\begin{tabular}{||c|c||}
\hline Molecule &Relative Abundance\\\hline
$^{229}$Th$^{+}$& 1\\\hline
$^{229}$Th(H$_{2}$O)$_{2}^{+}$& 0.55\\\hline	
$^{229}$ThO$^{+}$& 0.27 \\\hline	
$^{229}$Th(H$_{2}$O)$_{1}^{+}$&0.005  \\\hline	
$^{229}$ThO$_{2}^{+}$&$<$0.0001	\\\hline
\end{tabular}
 \caption{Molecular composition of the ions exiting the ion guide relative to the abundance of $^{229}$Th$^{+}$ measured using the JYFLTRAP Penning trap.}
	\label{tab:mols}
\end{table}

\noindent The most abundant molecule found was $^{229}$Th(H$_{2}$O)$_{2}^{+}$ which was approximately half the magnitude of $^{229}$Th$^{+}$ and arises due to water vapour being present in the system. $^{229}$ThO$^{+}$ is also formed in the ion guide through reactions with oxygen present on the surface of the source due to the electroplating process and by reactions with impurity oxygen molecules. This molecule is far more abundant than $^{229}$ThO$^{+}_{2}$ due to $^{229}$ThO$^{+}$ being more strongly bound and therefore surviving with a greater probability. Trace amounts of $^{229}$Th(H$_{2}$O)$_{1}^{+}$ are also formed in the ion guide and can again be attributed to water vapour in the system. $^{229}$Th(H$_{2}$O)$_{2}^{+}$ is seen in greater quantities than $^{229}$Th(H$_{2}$O)$_{1}^{+}$ due to metallic thorium favouring a higher coordination number. 

\noindent As the efficiency between the switchyard and Penning trap is dependent upon formation of molecules inside the rf-quadrupole cooler, this effect was studied. A $^{229}$Th$^{+}$ beam was injected into the cooler and the abundance of molecules exiting the device was measured. The results are tabulated (Table \ref{tab:coolermol}) and displayed relative to the abundance of surviving bare $^{229}$Th$^{+}$. The main molecule formed inside the cooler is $^{229}$ThO$^{+}$, primarily arising from oxygen present in the system after the breakup of water molecules in the rf-field.

\begin{table}[htbp]
	\centering
		\begin{tabular}{||c|c||}
\hline Molecule &Relative Abundance\\\hline
$^{229}$Th$^{+}$& 1\\\hline
$^{229}$ThO$^{+}$& 0.96  \\\hline	
$^{229}$ThO$_{2}^{+}$& 0.013	\\\hline	
$^{229}$Th(H$_{2}$O)$_{1}^{+}$& 0.01  \\\hline	
$^{229}$Th(H$_{2}$O)$_{2}^{+}$& $<$0.0001\\\hline	
\end{tabular}
 \caption{Molecular composition of ions exiting the rf-cooler relative to the abundance of surviving $^{229}$Th$^{+}$ after injection of a pure $^{229}$Th$^{+}$ beam. }
	\label{tab:coolermol}
\end{table} 

\begin{figure}
	\centering
		\includegraphics[width=\linewidth]{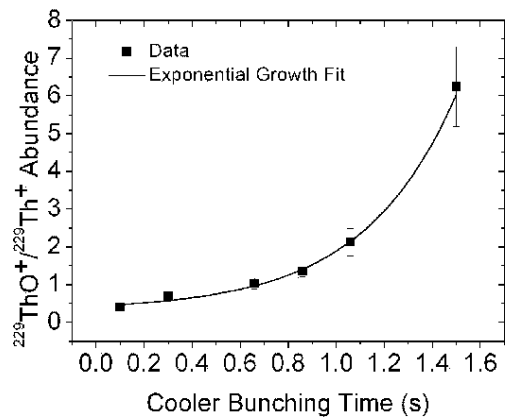}
	\caption{$^{229}$ThO$^{+}$/$^{229}$Th$^{+}$ ratio as a function of cooler bunching time normalised to the standard bunching time of 660 ms.}
	\label{fig:figure11}
\end{figure}

\noindent In an attempt to reduce the formation of $^{229}$ThO$^{+}$ inside the rf-cooler, the bunching time as a function of $^{229}$ThO$^{+}$/$^{229}$Th$^{+}$ ratio was investigated and is presented in Fig. \ref{fig:figure11}. An exponential growth fit to the data of the form  

\begin{equation}
y = y_{0} + A \text{exp}(x/t)
\end{equation}

was applied, whereby $y_{0}$ is the intrinsic switchyard to trap transmission efficiency and $t$ is a time constant for the formation of $^{229}$ThO$^{+}$ inside the cooler \cite{yuri}. 

\noindent This yielded an intrinsic switchyard to trap transmission efficiency of 32 $\pm$ 14\% and a molecular formation time constant of 0.39(7) s for $^{229}$ThO$^{+}$ production in the cooler.  This indicates that the impurity concentration level is 4.3 $\times$ 10$^{9}$cm$^{-3}$ assuming a reaction rate for production of $^{229}$ThO$^{+}$ identical to that of the reaction Th$^{+}$ + O$_{2} \rightarrow $ThO$^{+}$ + O \cite{johnsen}. This impurity concentration is comparable to the stated water contaminant abundance of the buffer gas. The possibility of molecular formation inside the purification trap was not studied here.\\

\noindent Accumulation of bunches on a timescale shorter than the reaction time constant inside the cooler improves the bare transmission efficiency of $^{229}$Th$^{+}$, but the intrinsic efficiency is unavoidable and is the main loss mechanism between the switchyard and experimental stations. Accounting for transmission losses between the ion guide and Penning trap, the overall $^{229}$Th$^{+}$ extraction efficiency is estimated at 0.06\% of the source strength. Missing efficiency arises through molecular formation inside the ion guide (Table \ref{tab:mols}) and due to an unknown fraction of neutral $^{229}$Th $\alpha$-recoils and molecules thereof. The neutral $^{229}$Th atomic fraction may be accessed via resonant laser ionisation \cite{iain} in the future.

\section{Conclusion}

\noindent The testing of differing electric field arrangements in this study has shown that a large potential gradient at large distances from the source of an electric field is best achieved through injection of electrons into a gas cell. This large potential difference directly contributes to both the ion guide efficiency for extraction of $\alpha$-recoils and reduction in extraction time of short lived ions. An ion guide efficiency for the extraction of $^{229}$Th$^{+} \alpha$-recoils is reduced compared to that of $^{221}$Fr$^{+}$, $^{219}$Rn$^{+}$ and $^{217}$At$^{+}$ and can be attributed to a combination of source geometry, chemical effects and an unknown neutral atom population. A $^{229}$Th$^{+}$ beam has been successfully extracted but impurity molecules restrict the bare ion fraction, most likely in favour of forming water and oxide compounds. Delivery of the ion beam to experimental stations is also reduced by the intrinsic transmission efficiency of the system, but formation of molecules inside the cooler can be almost eliminated by an appropriate choice of bunching time. A rate of 50 s$^{-1}$ $^{229}$Th ions at experimental stations has been achieved and a series of experiments using this newly available beam will follow. This will begin with collinear laser spectroscopy studies on the ground state and predicted low-lying isomeric state in $^{229}$Th. 

\section{Acknowledgements}
\noindent The authors would like to thank A. Nieminen for his contribution to the design of the ion guide. This work has been supported by the UK Engineering and Physical Sciences Research Council (EPSRC), the Natural Science and Engineering Research Council of Canada (NSERC), the Academy of Finland under the Finnish Centre of Excellence Program 2000-2005 and by the European Union Fifth Framework Programme "Improving Human Potential - Access to Research Infrastructure" Contract No. HPRI-CT-1999-00044. This work also benefits from the framework of agreement (project 8) between the Finnish and Russian Academies. BT is indebted to the Marie Curie Foundation for its financial support.

%\section{Results and Discussion}
%\label{results}

%\section{Conclusion and Outlook}
%\label{Conclusion}

%%%%%%%%%%%%%%%%%%%%%%%%%%%%%%%%%%%%%%%%%%%%%%%%%%%%%%%%%%%%%%%%%%%%%%%%%%%
 
\end{document}